\documentclass[aps,brb,twocolumn 
,Ndraft
,showpacs
,reprint
]{revtex4-1}

\usepackage{graphicx}
\usepackage{amsmath}
\usepackage{dcolumn}
\usepackage{multirow}
\usepackage{amssymb}
\usepackage{isomath}
\usepackage[usenames]{color}

\newcommand{\bl}{\vectorsym l}
\def\bd{\vectorsym{d}}
\def\br{\vectorsym{r}}
\def\bR{\vectorsym{R}}

\def\bV{\vectorsym{V}}

\def\bU{U}
\def\bpi{P}

\def\bp{\vectorsym{p}}
\def\dR{\vectorsym{{\delta}R}}
\def\bw{\vectorsym{w}}
\def\bW{\vectorsym{W}}

\newcommand{\beq}{\begin{eqnarray}}
\newcommand{\eeq}{\end{eqnarray}}

\begin{document}
\title{Metrics for measuring distances in configuration spaces}

\author{Ali Sadeghi,$^1$ S. Alireza Ghasemi,$^1$ Bastian Schaefer$^1$, Stephan Mohr,$^1$ Markus A. Lill$^2$  and Stefan Goedecker$^1$}
\affiliation{
$^1$Department of Physics, Universit\"{a}t Basel, Klingelbergstr. 82, 4056 Basel, Switzerland \\
$^2$Department of Medicinal Chemistry and Molecular Pharmacology, College of Pharmacy, Purdue University,
     575 Stadium Mall Drive, West Lafayette, Indiana 47907, United States}





\begin{abstract}
In order to characterize molecular structures we introduce configurational fingerprint vectors which are counterparts of 
quantities used experimentally to identify structures. 
The Euclidean distance between the configurational fingerprint vectors satisfies the 
properties of a metric and 
can therefore safely be used to measure dissimilarities between configurations in the high dimensional configuration space. 
We show that these metrics correlate well with the RMSD between two configurations if this RMSD is obtained from a global 
minimization over all translations, rotations and permutations of atomic indices. 
We introduce a Monte Carlo approach to obtain this global minimum of the RMSD between configurations. 
\end{abstract}


\maketitle

\section{Introduction}
Quantifying dissimilarities between molecular structures is an essential problem encountered in physics and chemistry.
Comparisons based on structural data obtained either from experiments or computer simulations
can help identifying or synthesising new molecules and crystals. 
Diversity analysis is at the heart of any 
structure prediction method in material science and solid state physics~\cite{OganovBook,minhop,amsler10,Neumann08,Oganov09}
and 
conformer search in  structural biology and drug discovery.~\cite{Downs,Velasquez06,Karakoc06,Kuntz,Geoffrey94,Zhang08,Gillet98}
In the latter case,  most of the proposed approaches~\cite{WhyMany,QSPR,Lemmen00}
use approximate methods that reduce the structure description information,
e.g. by excluding the side chains in a protein or a two dimensional representations of the molecule,~\cite{Allen00}
to speed up the searching procedure.~\cite{Schwarzer03}
In the case of solid state physics, 
fairly accurate dissimilarity measures are required. 
Within the structure prediction methods based on the evolutionary algorithms,~\cite{OganovBook}
the required diversity of populations can only be maintained if strongly similar configuration are eliminated. 
Within the Minima Hopping structure prediction method,~\cite{minhop} an identification 
of identical configurations is required as well to prevent trapping in funnels that do not contain the global minimum.
Some machine learning approaches~\cite{Rupp} are also based on similarity measures.

It is natural to characterize the dissimilarity between two structures $p$ and $q$ 
by a real number $d(p,q) \geq 0 $. 
In order to give meaningful results $d(p,q)$ should satisfy the properties 
of a metric, namely  
\begin{itemize}
\item  coincidence axiom: $  d(p,q)=0$  if and only if  $p\equiv q$,
\item  symmetry: $  d(p,q) = d(q,p)$,
\item  triangle inequality: $ d(p,q) + d(q,r)  \geq d(p,r) $.
\end{itemize}

The coincidence axiom  ensures 
that two configurations $p$ and $q$ are identical if their distance is zero,
and vice versa.
The triangle inequality is essential for clustering algorithms.
 If it is not satisfied, then it could happen that 
a configuration that belongs to one cluster in configuration space 
is also part of another cluster even though the distance between the two clusters   
is very large in the configuration space.

Since measuring distances between configurations is required in many applications,
a considerable effort has been made to find cheap, yet reliable, distance measures that
are not affected by the alignment of the two structures whose distance is being measured and by the 
indexing of the atoms in the structures. 
One class of approaches is based on a generalizations of standard 
physical descriptors such as coordination numbers. 
Cheng~\emph{et~al.}~\cite{Cheng04} used for instance the statistical properties (average, variance and bounds) of the coordination numbers 
while Lee~\emph{et~al.}~\cite{Lee03} used their weighted histograms 
in order to characterize the structures.  
Histogram-based methods were also used for the identification of crystalline structures.~\cite{Oganov10}
All these methods have several tuning parameters such as the width of histogram bins or cutoff radii for the determination 
of coordination numbers~\cite{Lee03} and their performance can critically depend on the choice  of these parameters. 

In this article we will introduce a family of parameter free metrics for measuring distances in configuration spaces.
We show that these metrics fulfil all the mathematical requirements and demonstrate their excellent performance for a representative set of benchmark systems including
covalent, metallic (simple or transition), ionic and organic structures.
The configurations in our test set are metastable low energy configurations obtained during a structure search  
using the Minima Hopping method~\cite{minhop}
on the density functional theory (DFT) level as implemented in the BigDFT code.~\cite{bigdft}

\section{RMSD} \label{sec:RMSD}
A configuration of $n$ alike atoms is uniquely represented 
by $\bR \equiv  (\br_1,\br_2, \dots, \br_n) \in \mathbb R ^{3\times n}$,
where  the column vector $\br_i$ 
represents the Cartesian coordinates of atom $i$.
A distance based on the naive Frobenius norm
\beq
\|\bR^p-\bR^q\|= \Big( \sum_{i=1}^{n_{}}\|\br_i^p-\br_i^q\|^2\Big)^{1/2} \label{eRMSD}
\eeq
can not be used to compare two configurations p  and q, because  
it is not invariant with respect to translations or rotations of one configuration relative to the other.
For this reason the commonly used  root-mean-square distance (RMSD)
is defined as the minimum Frobenius distance  over all translations  and rotations. 
By minimizing $\sum_i^n\|\br_i^p+\vectorsym{d} - \br_i^q\|^2 $
with respect to the translation $\vectorsym d$
one obtains 
$\sum_i^n (\br_i^p+\vectorsym{d} - \br_i^q )=0 $, 
i.e. the required translation is the  difference  between the centroids
$\vectorsym{d} = \frac 1 n \sum_i^n \br_i^{q} - \frac 1 n \sum_i^n \br_i^{p}.$
Therefore we will assume in the following that all $\br_{i}$ are measured with respect to the centroids of the 
corresponding configuration 
which allows us to drop the minimization with respect to the translation $\bd$.
Then, finding the rotation $\bU$ around the common centroid which minimizes
\beq \label{eq:RMSDl}
 RMSD_l(p,q)=\frac 1 {\sqrt n} \min_{\bU} \|\bR^p-\bU \bR^q \| 
\eeq
is a local minimization problem 
and hence we denote this version of the RMSD by RMSD$_l$.
The Kabsch algorithm~\cite{Kabsch} provides the solution to this problem based on the Euler angles. 
Like many others, we perform the local minimization by an alternative method based on quaternions~\cite{Horn}
(see Appendix~\ref{apx:Q})
 which is more stable and numerically very cheap.~\cite{dill,QRMSD}

The RMSD$_l$ is however not invariant under index permutations of chemically identical atoms. 
If the configuration $p$ and $q$ are identical, 
Eq.~(\ref{eq:RMSDl}) will be different from zero if we permute for instance in $\bR^q$ 
the positions $\br_i^q$ and $\br_j^q$ of atoms $i$ and $j$.
The minimum Frobenius distance obtained by considering all possible index permutations 
for an arbitrary rotation $\bU$ is   
\beq \label{eq:RMSDp} 
 RMSD_\bpi(p,q)=
 \frac 1 {\sqrt n}
 \min_{\bpi}  \|\bR^p- \bU \bR^q  \bpi\|
,\eeq
$\bpi$ being an $n\times n$ permutation matrix.
This assignment problem is solved in polynomial time 
using the Hungarian algorithm.~\cite{Hungarian} 

What is really needed is a solution of the combined problem
of the global minimization over all 
rotations and permutations, namely
\beq \label{eq:RMSD} 
 RMSD(p,q)=
 \frac 1 {\sqrt n}
 \min_{\bpi,\, \bU}  \|\bR^p- \bU \bR^q\bpi \|
.\eeq
The  global minimum  RMSD 
fulfills all the properties of a metric. 
The coincidence and symmetry properties are easy to see.
Using the standard triangle inequality, 
the proof of the triangle property is as follows:
\beq 
&& RMSD(p,q)+RMSD(q,r)
\nonumber \\
&&=\frac 1 {\sqrt n}  \min_{\bpi,\, \bU} \|\bU \bR^p\bpi- \bR^q\| + \frac 1 {\sqrt n} \ \min_{\bpi,\, \bU} \|\bR^q-\bU \bR^r\bpi\| 
\nonumber \\
&&=\frac 1 {\sqrt n} \|\bU_{pq}\bR^p\bpi_{pq}  - \bR^q\| + \frac 1 {\sqrt n} \|\bR^q-\bU_{rq} \bR^r\bpi_{rq}\| 
\nonumber \\
&&\geq   \frac 1 {\sqrt n}\|\bU_{pq}\bR^p\bpi_{pq} - \bR^q + \bR^q -\bU_{rq}\bR^r\bpi_{rq} \| 
\nonumber \\
&&\geq   \frac 1 {\sqrt n}\|\bR^p -\bU_{rp} \bR^r\bpi_{rp}\| 
\nonumber \\
&&= RMSD(p,r)
\nonumber 
\eeq
where $\displaystyle\min_{\bpi,\, \bU} \|\bU \bR^p\bpi- \bR^q\|$ is shown by 
$ \|\bU_{pq}\bR^p\bpi_{pq}  - \bR^q\|$ for convenience.

Since $\bU$ and $\bpi$ are not independent, 
no algorithm exists which can find the global RMSD within polynomial time.
Just doing a search by alternating rotation and permutation steps
using local  minimizations and the Hungarian algorithm, respectively, 
is not guaranteed to converge to the global minimum with a finite number of steps.
Trying out all possible permutations would lead to a factorial increase of the computing time with respect to $n$ 
and this approach is therefore not feasible except for very small systems. 
In some applications, one might apply restrictions into the permutations in order to reduce the size of the permutation space. 
For instance, in an application to organic molecules only equivalent atoms has to be permuted, 
e.g. see Ref.~\cite{Wales12}.
Equivalent atoms in an organic molecule are considered for example those that have identical connectives 
 determined by the Morgan algorithm.~\cite{Morgan,Morgan2}
 For all kind of molecular structures, however, such a grouping   of identical atoms ones is not possible.

\begin{figure}[h!] \begin{center}
\includegraphics[width=.49\textwidth]{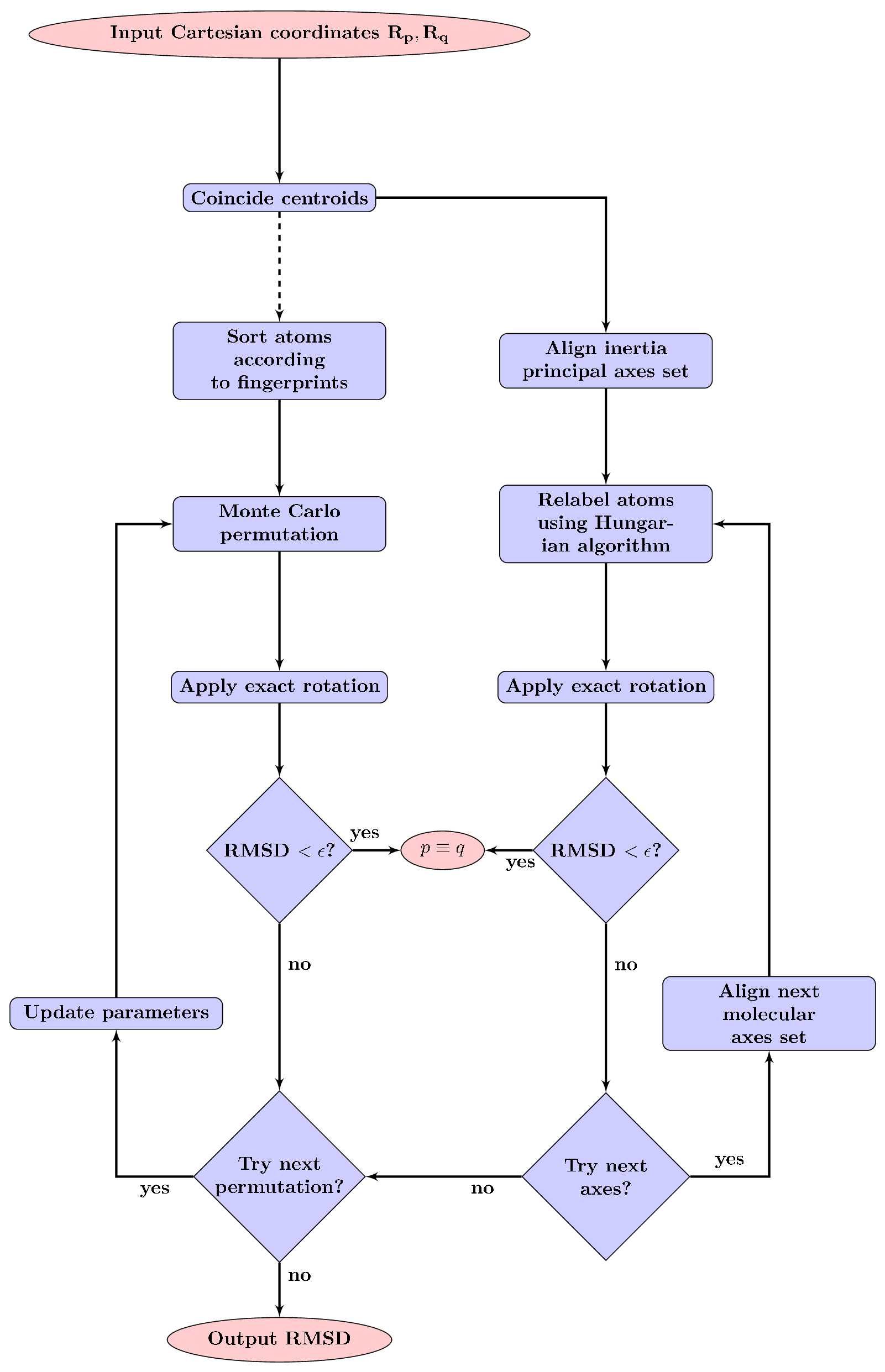}
\caption {\label{fig:flowchart}  
Flowchart of the algorithm of global minimization of RMSD 
in two major steps.
The loop on the right runs over several sets of axes  and 
matches atoms of a pair of configurations via  aligning their molecular axes.
The left loop shows the Monte Carlo (MC) permutation of identical particles
while the parameters are dynamically tuned to obtain an acceptance rate close to 50\%. 
The dashed line means that the right loop can be excluded.
}\end{center} \end{figure} 

\begin{figure} \begin{center}
\includegraphics[angle=-90,width=.42\textwidth]{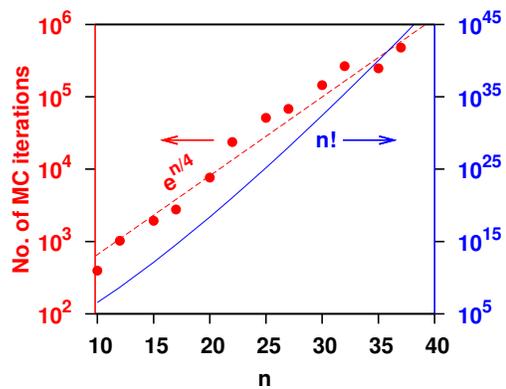}
\caption {\label{fig:MC} 
Average number of the MC iterations required to obtain the global RMSD between randomized LJ clusters
as a function of the number of particles $n$. 
The dashed line ($41\exp(n/4)$) is obtained by least square fit.
For comparison, $n!$ is also plotted with solid line. 
}
\end{center} \end{figure}

We use a two-stage  method for finding the global  RMSD with moderate computational effort.
 The flowchart of the algorithm is depicted in Fig.~\ref{fig:flowchart}
with the two different stages shown on the left and right sides.
In the first stage we try to find the optimal global alignment of the two structures being compared.
We first align two of the three principal axes of inertia of one configurations with the corresponding axes of the other one. 
A trial alignment is always followed by the application of the Hungarian algorithm to find the index permutation that gives the smallest RMSD.~\cite{Helmich12}
The index matching in the Hungarian algorithm is done in the Cartesian space by associating to each atom $i \in$~p 
the closest atom $j \in$~q
such that $\sum_i^n\|\br_i^p-\br_j^q\|$ is minimal. In other words, the columns of the $n\times n$ matrix made by $\|\br_i^p-\br_j^q\|$ 
are reordered 
such that its trace is minimal.
The implementation of the Hungarian algorithm based on Ref.~\cite{Hungariancode} finds 
the optimal index permutation within polynomial time and with a small prefactor.
After this initial index matching, 
a rotation using quaternions is applied to refine the molecular alignment. 
If the required rotation is significant, the atomic index assignment should be repeated.
This whole procedure is iterated until the atomic indices remain fixed after applying the rotation. 
This procedure has allowed us to detect all identical configuration in this first stage, 
as seen  in Table~\ref{tab:}.

Since all the global alignment methods are empirical and can fail we apply several of them successively. 
After the first global alignment based on the principal axes of inertia we apply some more alignments steps 
based on axes which are derived from local atomic fingerprints (see next section). 
We set up an overlap matrix with s and p type Gaussian orbitals (see Appendix~\ref{apx:OM}) and find its principal eigenvector 
(i.e. the eigenvector with the largest eigenvalue; see Fig.~\ref{fig:atomicfp}). 
Defining  $\bw_i=s_i\bp_i$, where $s_i$ and $\bp_i$ are respectively s- and p-type components of the principal eigenvector 
belonging to atom $i$ we can form two axes  $\bW$ and $\bW'$
\beq \label{eq:W} 
\bW &=& \sum_i^n \bw_i , \\ \bW' &=& \sum_i^n \bw_i \times \br_i\label{eq:W'} 
\eeq 
where the sum runs over the atoms,
$\br_i$ represents the positions of atoms with respect to the center of mass
 and $\times$ denotes the cross product.
First, we align $\bW^q$ with  $\bW^p$ and then rotate q around it such that 
the plane made by ($\bW^q$,$\bW'^q$) coincides with the plane made by ($\bW^p$,$\bW'^p$).
Depending on the width  of the Gaussian used to construct the overlap matrix, several sets of 
axis may be constructed and tried one-by-one in this stage.
If the alignment according to a new set of axes results in a smaller RMSD, we accept it.
In Table~\ref{tab:} we show the results of the alignment of the principal axes of inertia 
as well as three sets of $(\bW,\bW')$ axes obtained by three different Gaussian widths $\alpha$.

If a small enough  RMSD is not found, we
enter into an iterative stage   (see left side of Fig.~\ref{fig:flowchart}) 
 where randomly chosen atoms are permuted
 within a thresholding Monte Carlo (MC) approach
followed by applying the optimal rotation.
The iteration stops when the global minimum RMSD does not decrease any more.

\begin{table*}
\begin{center}
\caption{\label{tab:} 
Number of remaining distinct configurations, average RMSD 
and average CPU-time (on single 2.4~GHz Intel core) for superimposing one pair of configurations
at different steps of the two-stage RMSD global minimization.
In the first stage, the principal axes of inertia as well as
three molecular sets of axes obtained from vectorial atomic fingerprints are used.
Every molecular alignment is always followed by the application of the Hungarian algorithm  to find the optimal index permutation.
In the second stage, random permutations are tried out which are followed by local minimization to get the optimal rotation.
Because of the stochastic nature of the MC part, the reported values might change in different runs. 
}
 \begin{tabular}
{  l c | c  c c| c  c  c | c c  c}
\toprule
\multicolumn{2}{c|}{ }  &\multicolumn{3}{c|}{ Si$_{32}$ } &  \multicolumn{3}{c|}{ Mg$_{26}$ } &  \multicolumn{3}{c}{ C$_{22}$H$_{24}$N$_2$O$_3$} \\
      &       & remaining  & $\overline{\text {RMSD}}$ &  $\overline{\text t}_\text{CPU}$ & remaining & $\overline{\text {RMSD}}$& $\overline{\text t}_\text{CPU}$ & remaining  & $\overline{\text {RMSD}}$ &  $\overline{\text t}_\text{CPU}$  \\ 
      &       & distinct  & [\AA] & [s] & distinct   & [\AA] & [s] & distinct   & [\AA] & [s]  \\ 
 \hline
\multicolumn{2}{c|}{Unanalyzed}&317 & 1.40  &     &111 & 3.44 & & 60 & 2.75 \\ 
 \hline
\multirow{4}{*} {Axes Alignment} &
       axes of inertia         &184 & 1.16  &          & 60 & 1.08  &          & 42 & 1.93 &              
 \\&  $(\bW,\bW')_{\alpha_1}$      &184 & 1.06  &          & 59 & 1.06  &          & 42 & 1.89 &           
 \\&  $(\bW,\bW')_{\alpha_2}$      &184 & 1.04  &          & 59 & 1.03  &          & 42 & 1.81 &           
 \\&  $(\bW,\bW')_{\alpha_3}$      &184 & 1.02  &$<0.001$& 59 & 1.01  &$<0.001$& 42 & 1.78 &$<0.001$ 
 \\   \hline                                                                                          
\multirow{5}{*} {Monte Carlo} &                                                                       
      iter.=$ 10^{3}$          &184 & .978  & 0.03   & 59 & .985  & 0.02   & 42 & 1.52 & 0.05       
\\  & iter.=$ 10^{4}$          &184 & .910  & 0.13   & 59 & .864  & 0.11    & 42 & 1.51 & 0.13       
\\  & iter.=$ 10^{5}$          &184 & .852  & 1.1      & 59 & .852  & 1.0    & 42 & 1.51 & 1.6         
\\  & iter.=$ 10^{6}$          &184 & .792  & 12.1     & 59 & .824  & 10    & 42 & 1.51 & 15 
\\  & iter.=$ 10^{7}$          &184 & .791  & 132      & 59 & .824  & 119      & 42 & 1.51 & 163 
 \\  \hline     \hline 
 \end{tabular}
 \end{center}
 \end{table*}

As seen in Table~\ref{tab:}, the number of required MC iterations depends on the system size.
For instance, for the biomolecule $10^4$ MC iterations
(which take on average 0.13 second on a single 2.4~GHz Intel core)
are sufficient to find the global minimum RMSD between two configurations of this molecule.
For a more systematic investigation of the scaling, we take the global minima of the Lenard-Jones (LJ) clusters with different sizes
 and apply random displacements of the unit magnitude to every  atom 
(i.e. the RMSD between the randomized structures is almost one in the LJ length units). 
The averaged number of required MC iterations
to get the asymptotic value of the RMSD (as obtained by $10^7$ iterations), 
as a function of the cluster size $n$ is shown in Fig.~\ref{fig:MC}.
Even though the number of iterations increases exponentially it is several orders of magnitude smaller than the number of possible 
permutations, i.e. $n!$. 

\section{Fingerprint Distances as metrics}
While the RMSD can be considered as the  most basic quantity to measure the dissimilarities,
finding the global minimum RMSD is numerically costly.
Only in case that two structures are nearly identical the global minimum of RMSD is calculated with a polynomial computational time 
because no MC permutation is then required. 
Otherwise, even if the above described algorithm is used, the computational time increases exponentially with the number of permutable particles. 
In the following  we will therefore 
introduce  a family of 
metrics which are cheaper to calculate than the global RMSD 
yet in good agreement with it.
We consider symmetric $N \times N$ matrices 
whose elements depend only on the interatomic distances $r_{ij}=\|{\bf r}_i - {\bf r}_j\|$ 
of an $n$-atom configuration.
Vectors $\bV$ containing eigenvalues of such a matrix form a configurational fingerprint which allows to identify a structure. 
The normalized Euclidean distance 
\beq \label{eq:dV}
\Delta_{\bV}(p,q)=\frac 1 {\sqrt N} \|\bV^p-\bV^q\|
\eeq
measures the dissimilarly between p and q with no need to superimpose them.
For a vector 
 whose elements are formed from the elements of selected eigenvectors, each element can 
be associated to an atom and the ensemble belonging to one atom forms 
an \emph{atomic} fingerprint or descriptor of the local environment of the atom. 
However, such eigenvectors are not used in this work for describing the whole structure;
we use them only in Eqs.~(\ref{eq:W}) and (\ref{eq:W'}) to attribute individual atoms. 
Since the matrix depends only on interatomic distances, the 
same holds true for the eigenvalues and eigenvectors, and $\bV$ 
is thus invariant under translations, rotations and reflections of the configuration.
In order to make  $\Delta_\bV$ also independent of the atomic indices, the elements of each $\bV$ 
are sorted in an ascending order. This sorting can introduce discontinuities in the first derivative 
of the fingerprint distance with respect to changes in the atomic coordinates (e.g. when there is a crossing 
of eigenvalues) but does not destroy the important continuity of the fingerprint distance itself. 

The coincidence axiom for a configurational fingerprint is satisfied if the dimension $N$ of the matrix is sufficiently 
large and if therefore the resulting fingerprint vector is sufficiently long.
To see this, 
let us consider two configurations $p$ and $q$ which are close. 
The difference of the fingerprint vectors is then given by a first order Taylor expansion
\begin{equation} \label{LS} 
   \bV^p-\bV^q \simeq D (q) (\bR^p - \bR^q), 
\end{equation}
Note that, instead of the $3\times n$ matrix notation used in Sec.~\ref{sec:RMSD}, 
hereafter we use  a column vector $\bR \in \mathbb R ^{3n}$ for representing the atomic coordinates.
Since  $\bV$ is a column vector of length $N$,
the first derivative $D(q)\equiv \frac{ \partial \bV}{\partial \bR} \big|_{\bR=\bR^q}$ 
is a $N\times 3n$ matrix.
We assume that $D$ has always the largest possible rank for the three types of matrices discussed in more detail in this Section. 
For the Hamiltonian matrix this maximal rank  $r_\text{max} $ equals $\min(N,3 n-6)$ 
if all $N$ eigenstates included in the fingerprint vector are bound. 
For the overlap matrix $r_\text{max} $ equals $ \min(N-1,3 n-6)$ because the diagonal elements 
are independent of the configuration. 
For the Hessian matrix $r_\text{max}=3n-6$ for configurations that are local minima with respect 
to the interaction potential and $r_\text{max}=3n-3$ for all other cases.~\cite{mfield}

\begin{figure} \begin{center}
\includegraphics[angle=-0 ,width=.25\textwidth]{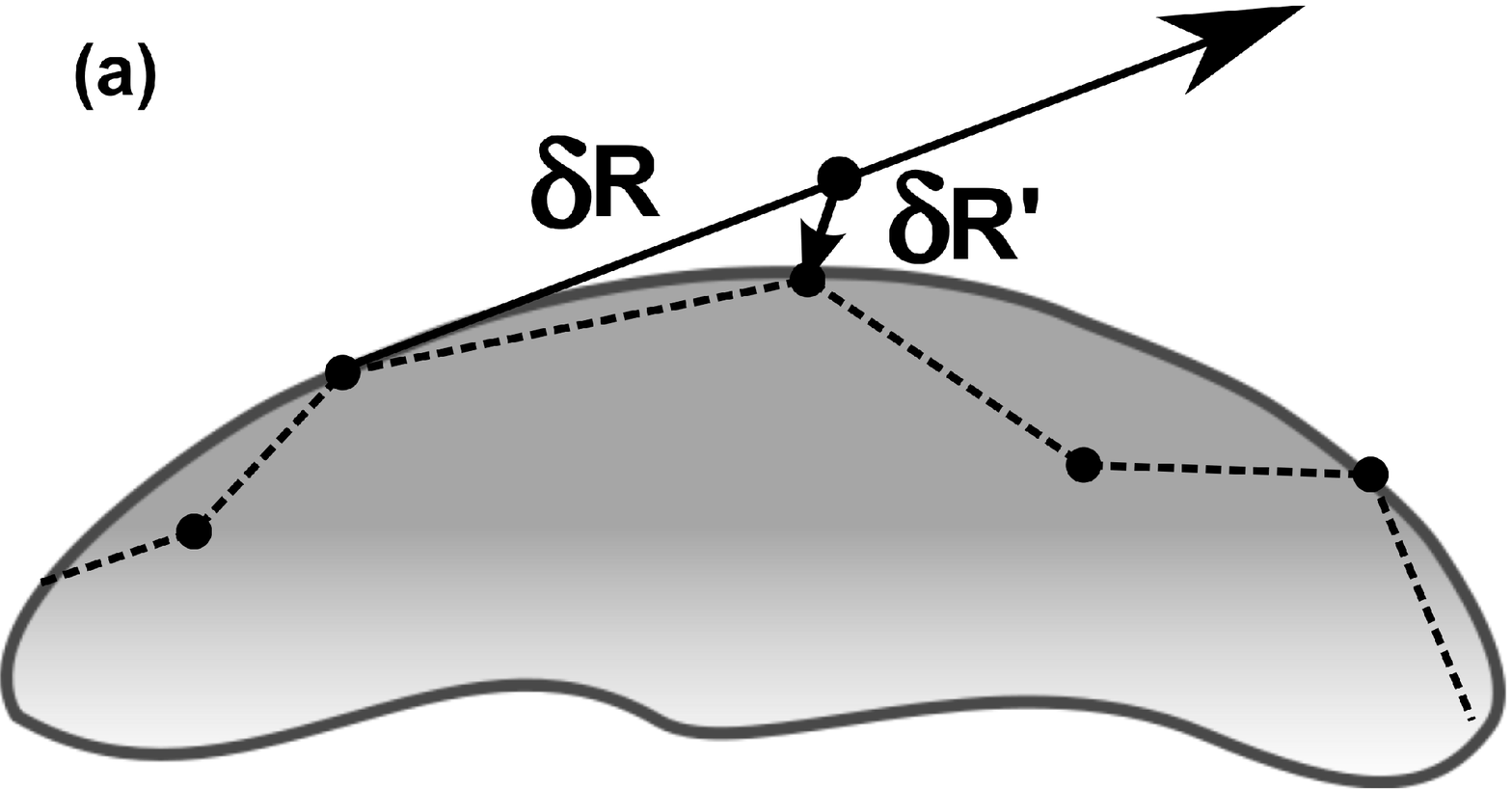} \\
\includegraphics[angle=-90,width=.10\textwidth]{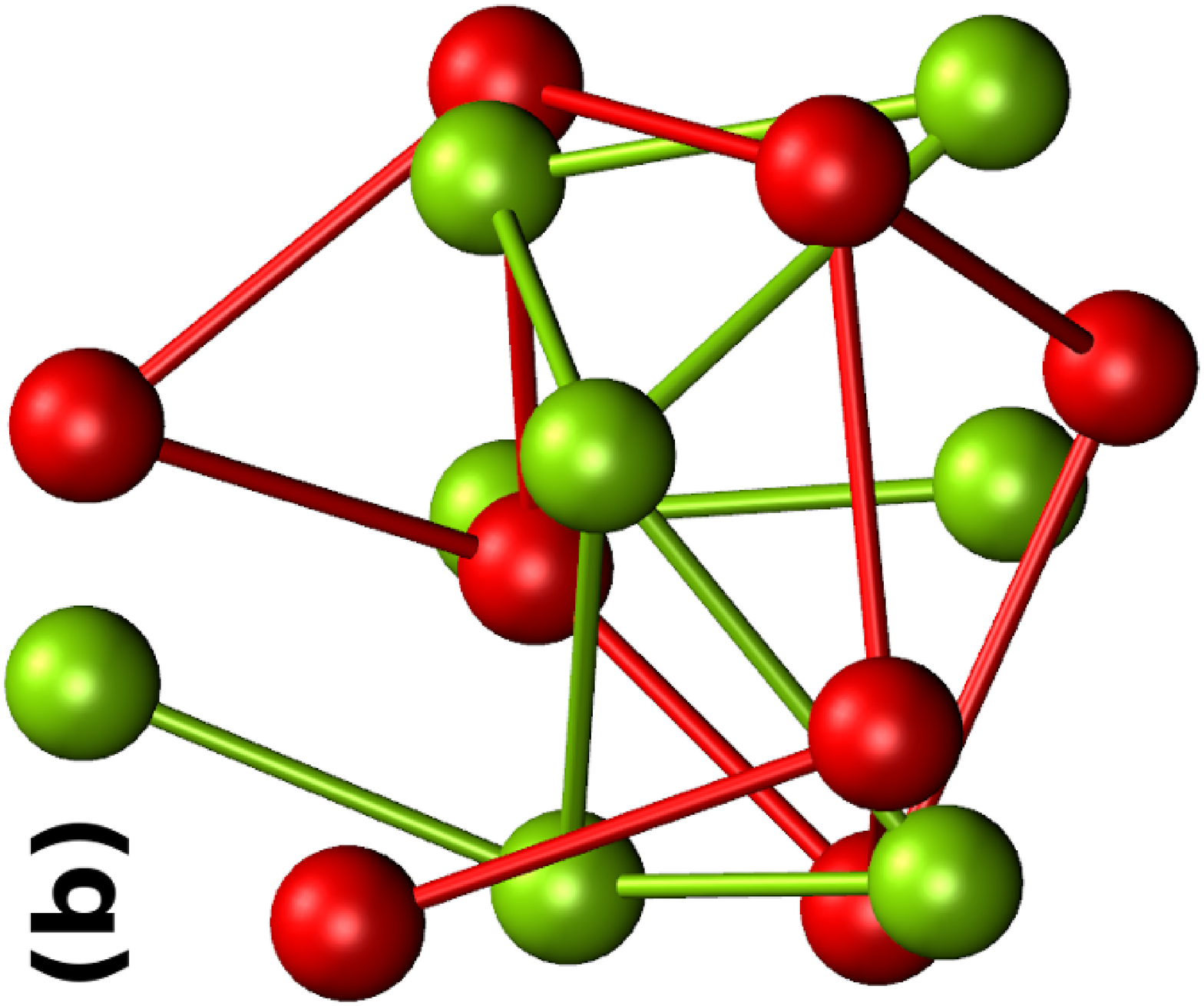}
\includegraphics[angle=-90,width=.37\textwidth]{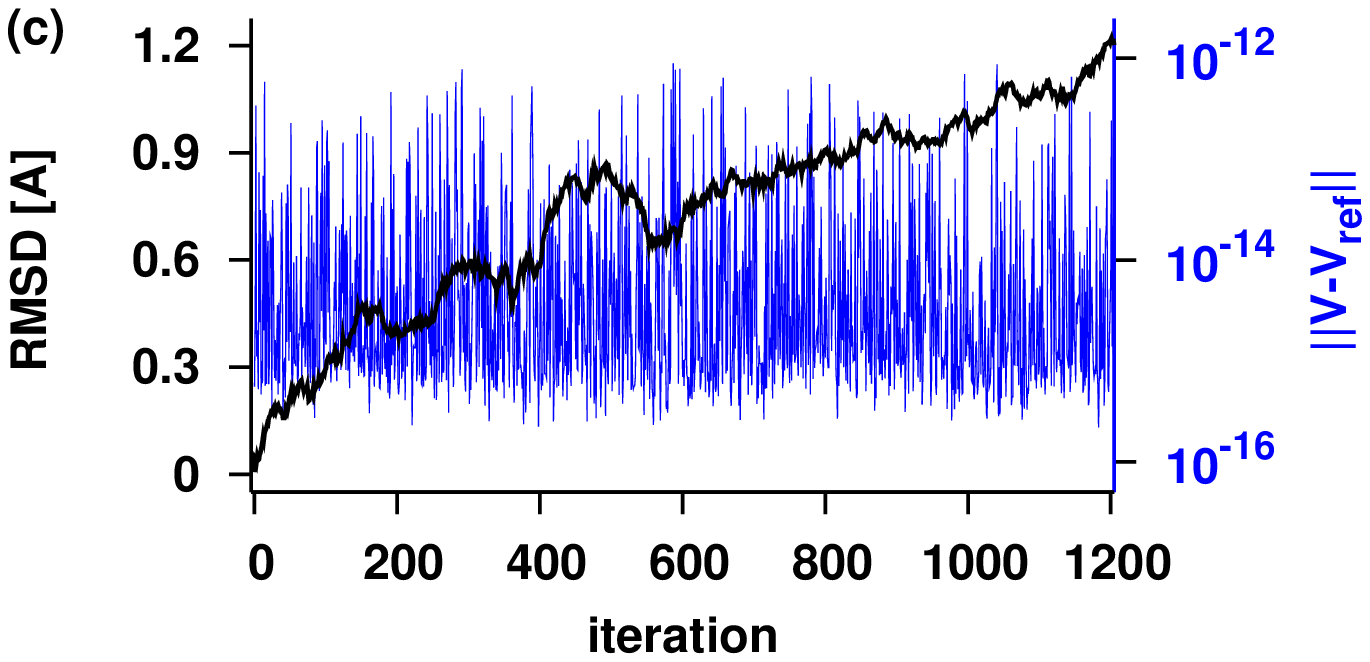}
\includegraphics[angle=-90,width=.105\textwidth]{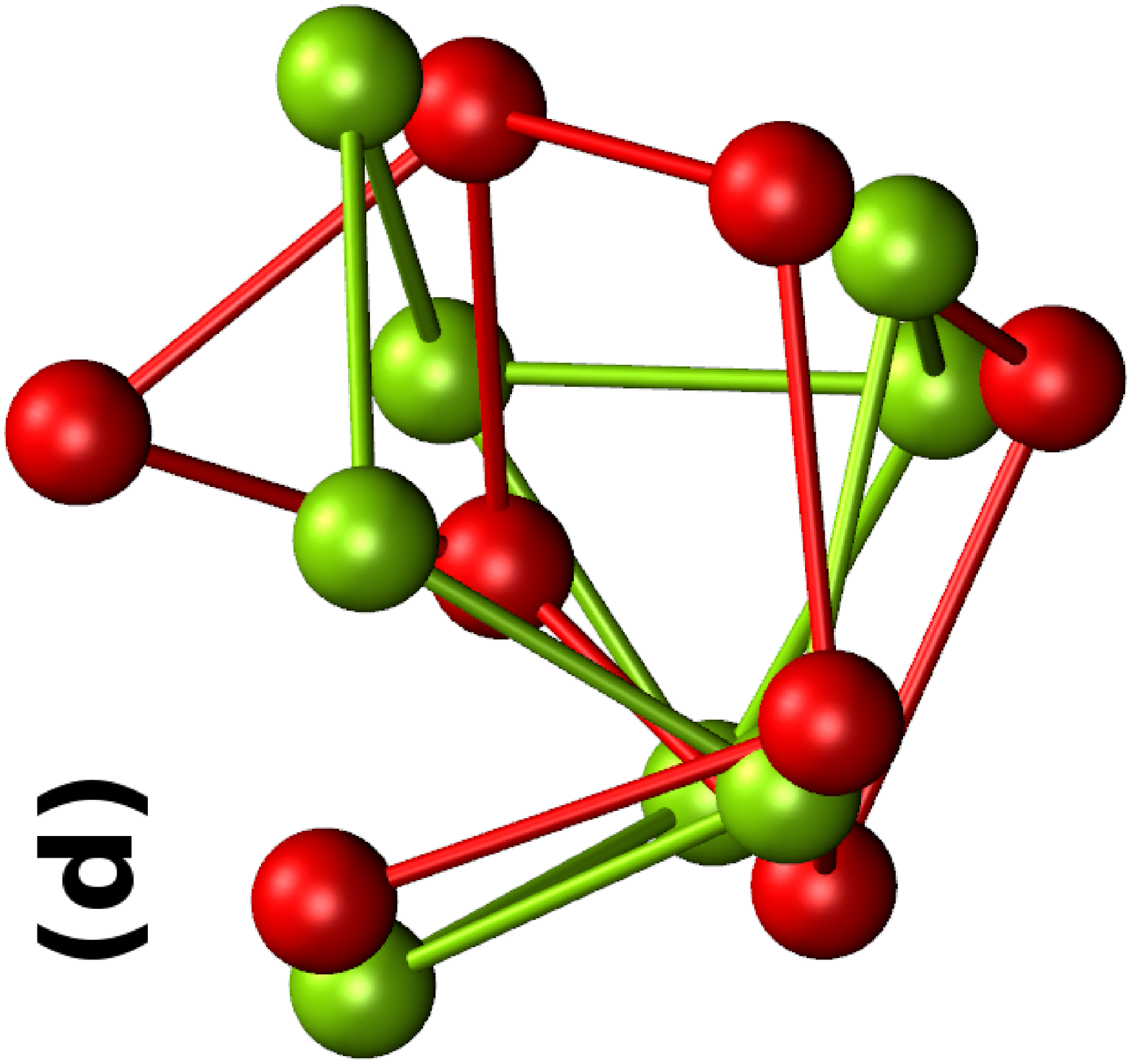}
\includegraphics[angle=-90,width=.37\textwidth]{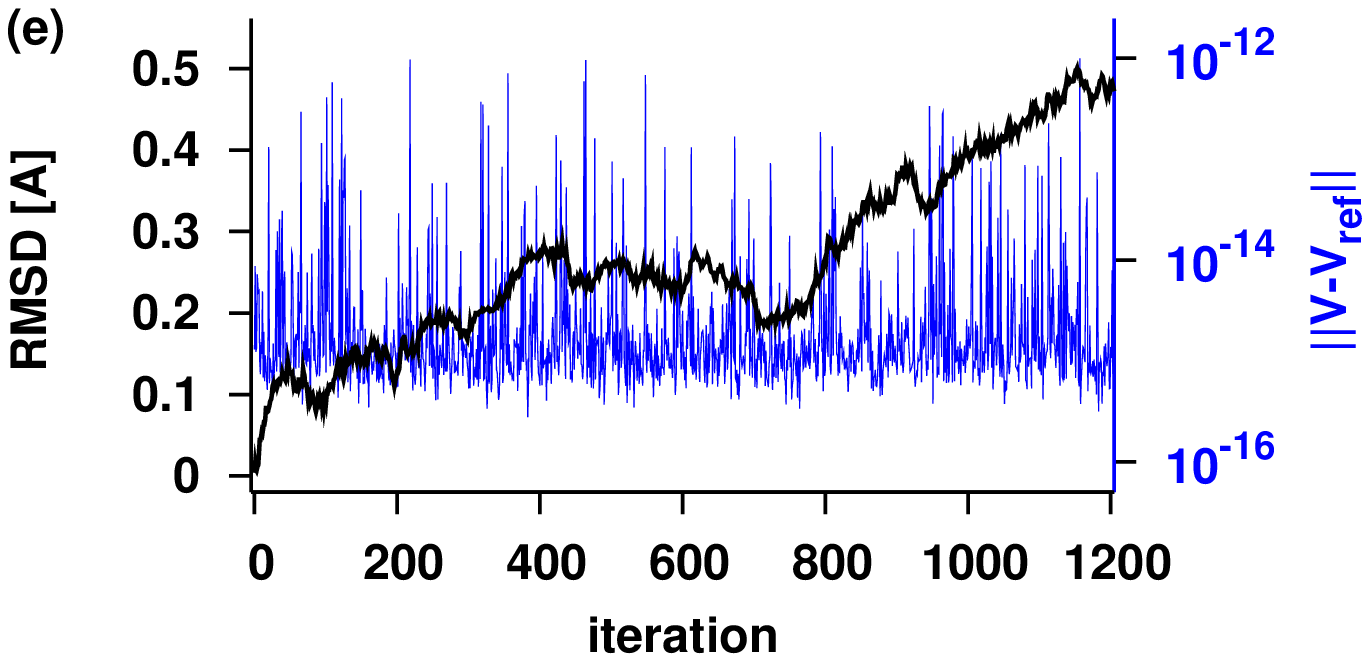}
\caption {\label{fig:nullspace} 
(a) Schematic illustration of the exploration of the hypersurface 
defined by $\bV=\bV_\text{ref}$
consisting of iterative movements along $\dR$ (in the null space of $D$)
 followed by Newton step(s) $\dR'$ to come back to the hypersurface.
Panel (b) shows two configurations (in red and green) of a Si$_{8}$ 
cluster whose fingerprint vectors of length $n$, obtained from an overlap matrix 
with one set of s-type GTO's, are identical. Panel (c) shows the evolution of the RMSD 
during the exploration of the hypersurface leading from the red structure to the 
green structure.
Panels (d) and (e) contain the some information as panels (b) and (c) but 
for a fingerprint of
length $2n$ obtained from an overlap matrix with two  sets of s-type GTO's.
In both cases $\|\bV-\bV_\text{ref}\|$ is vanishingly small.
}
\end{center} \end{figure}

If $r_\text{max}$ is less than $3 n-6$ one can find on a hypersurface of dimension $ 3 n-6 -r_\text{max} $ 
(i.e. the nullity of $D$) 
configurations with identical fingerprint vectors, which are given as a solution of the equation 
\begin{equation} \label{DdR} 
 D \dR  = \bf 0 .
\end{equation}
Formulated in words, configurational displacement vectors $\dR$ which are in the null space of $D$ leave the fingerprint invariant to first order.
For configurations  
which are further apart the  first order approximation breaks 
down but Eq.~(\ref{DdR}) can still be used as a starting point for mapping out such a hypersurface iteratively. 
We perform a move with a small 
amplitude along a vector $\dR$ in the null space of $D$. 
To correct for the small second and higher order deviations of the eigenvalues away from the 
hypersurface of constant eigenvalues defined as $\bV=\bV_\text{ref}$ we then solve 
\beq \label{DdR'}
D\dR'=\bV_\text{ref} - \bV
\eeq
for the required displacement $\dR'$. 
Like Eq.~(\ref{LS}), the latter equation does not have a unique solution and we can therefore choose an arbitrary set of $r_\text{max}$ coordinates which we want to modify in order to go back onto the 
hypersurface of constant eigenvalues.
If the corresponding $r_\text{max}\times r_\text{max}$ matrix made out of $D$ was ill-conditioned, we try out another set
of $r_\text{max}$ atomic coordinates to modify to ensure that Eq.~(\ref{DdR'}) is solved accurately. 
 Since this going back to the hypersurface requires only tiny displacements a single solution of the linear system is sufficient.
 If this was not the case it could be repeated which would correspond to a Newton iteration. 
By iterating this procedure of moves along the null space 
followed by moves that bring us exactly back on the hypersurface 
we can obtain clearly distinct 
configurations whose fingerprints are identical up to machine precision. 
Such examples are shown in Fig.~\ref{fig:nullspace} 
 where the procedure is also illustrated schematically. 
Note that at each iteration we orthogonalize $\dR$ of the previous iteration
to the row space of current $D$.
This reduces the probability of coming backward to the starting point,
as is featured in the diffusion-like  pattern of RMSD versus iteration.

The constructive iterative procedure outlined above shows how a hypersurface of constant fingerprint can be constructed if the length of the fingerprint is short.
What we would like to show however is the opposite, namely that no distinct configurations with identical fingerprints exist if the fingerprint is long enough.
Since the fingerprint distance is
a non-linear function, it can in principle  not be excluded that two distinct configurations with identical
fingerprints exist even if the fingerprint vector is longer than the threshold value. Since we recommend for a unique identification fingerprints 
which are considerably longer than the threshold value, namely fingerprints of length 
$3n$ or even $4n$ it is however extremely unlikely that such configurations exist and the coincidence axiom can be taken to be fulfilled. 
To confirm this assumption numerically as well, we did extensive numerical searches where we tried to find a second configuration which has a fingerprint which is 
identical to the fingerprint of a reference configuration. The initial guess for the second configuration was random and then this second configuration was 
moved in such a way as to minimize the difference between the fingerprints. All these numerical minimizations lead to 
non-zero local minima, i.e. we were not able to find numerically any violation of the coincidence axiom for  
vectors of length $3n-3$ based on the Hessian matrix and vectors of length $4n$ based on an overlap matrix with s and p orbitals.

\begin{figure} \begin{center}
\includegraphics[width=.4\textwidth]{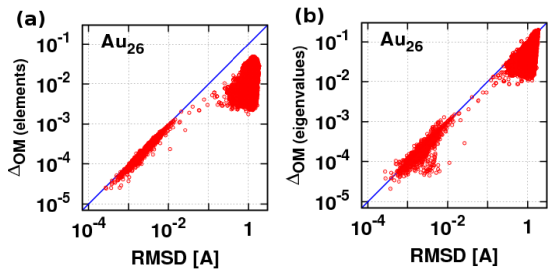}
\caption {\label{fig:elementsVSevals}   
Correlation of the pairwise Euclidean distances based on vectors consisting either of all of the 
sorted elements of the overlap matrix (a) or eigenvalues of this matrix (b) and the RMSD 
for 1000 metastable configurations of a 26 atom gold cluster. The gap in the fingerprint distances 
between identical and distinct configuration is larger if eigenvalues are used (panel a). 
}
\end{center} \end{figure}

Even though the eigenvalue vector is much shorter than the vector containing all matrix elements,
the fingerprint distances based on the eigenvalues  
are better than those obtained by sorting all the matrix elements depending on interatomic distances into a vector. 
One can in some cases construct distinct so-called homometric 
configurations~\cite{homometric} for which the fingerprint vectors of the sorted matrix elements are identical 
whereas the eigenvalue vectors are not identical and allow thus to distinguish between them. 
In addition, our empirical results of Fig.~\ref{fig:elementsVSevals} show that the gap 
between  identical and distinct pairs is larger for the eigenvalues 
than for the sorted matrix elements. 
  Because the geometry relaxations were stopped when the force on each
  atom is within 0.01~eV/\AA, identical configurations are in practice
  identical only up to some finite precision. Two configurations are considered
  to be identical if their distance is below
  a certain threshold. An unambiguous  threshold for distinguishing between distinct and non-distinct 
  configurations can only be found if a well detectable gap
  exists in the distance space. Hence the existence of a large gap is an important benefit of a fingerprint method. 

\begin{figure} \begin{center}
\includegraphics[angle=-0, width=.45\textwidth]{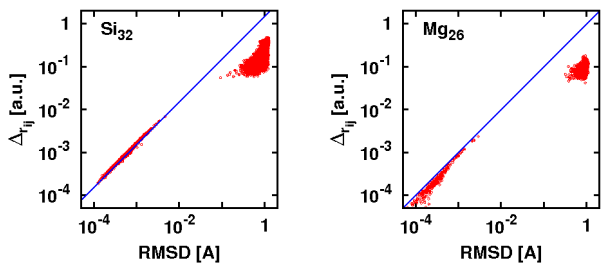}
\caption {\label{fig:rij}   
Correlation of Euclidean distance of the sorted interatomic distances with RMSD 
for the metastable configurations of the Si$_{32}$ and Mg$_{26}$ clusters. The gap 
that allows to discriminate distinct from non-distinct configurations is smaller 
in both cases compared to the fingerprints based on eigenvalues. 
}
\end{center} \end{figure}

\begin{figure*} \begin{center}
\includegraphics[angle=0, width=\textwidth]{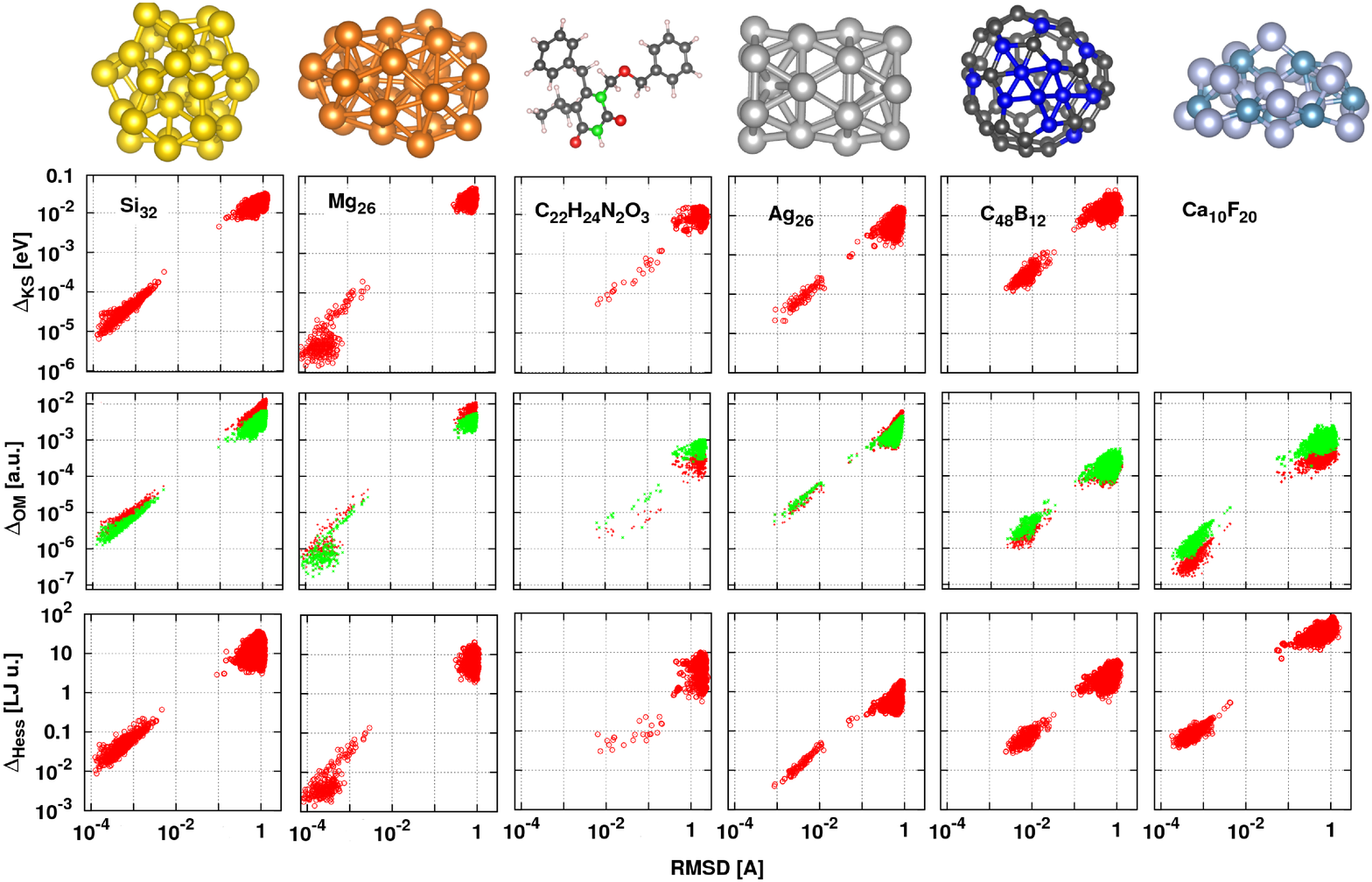}
\caption {\label{fig:compare} 
 Correlation of metrics based on the eigenvalues of the Kohn-Sham Hamiltonian matrix (first row),
the overlap matrix (second row) and the Lennard-Jones Hessian matrix with the RMSD
for sets of  
semiconductor (silicon), simple metal (magnesium), organic (6-benzyl-1-benzyloxymethyl-5-isopropyl uracil), 
transition metal (silver), covalent fullerene-type (C$_{48}$B$_{12}$) and ionic (calcium fluoride) clusters. 
Shown on top are representative configurations.
Each set consists of few hundred configurations, all being low-energy local minima within DFT, except those of Ca$_{10}$F$_{20}$
which are local minima of the Tosi-Fumi potential (parameters from Ref.~\cite{Benson}).
For the other five sets from left to right
we used, respectively,  
$64 = 2n$, $26 = n$, $n<70 < 2n$, $26 = n$ and $n<114 < 2n$
 number of Kohn-Sham eigenvalues corresponding to the occupied, valance states,
$n$ being the number of atoms.
For the overlap matrix, results for both s-only (red) and s-and-p (green) overlap matrices are shown, leading to fingerprint vectors of lengths $n$
and $4n$, respectively. For the Hessian matrix $3n-3$ eigenvalues are non-zero.
Even when the length of the fingerprint vector is not longer than $3n-6$ and hence the coincidence axiom is not satisfied,
the agreement with the RMSD is perfect.
}
\end{center} \end{figure*}

In an application to Ni clusters 
Grigoryan~\emph{et al.}~\cite{Grigoryan03} 
used the sorted interatomic distances 
to find the similarities between an $(n-1)$-atom cluster and $(n-1)$-atom parts of an $n$-atom cluster.
This similarity measure also leads to a gap which is smaller than the one obtained from 
eigenvalue based fingerprints of either the corresponding  $r_{ij}$ matrix or the matrices proposed in 
this article (cf. Figs.~\ref{fig:rij} and \ref{fig:compare}). So it seems to be a general feature that fingerprints based on the 
eigenvalues are better than those based on sorted matrix elements.

In the following we will describe several matrix constructions which can be used for fingerprinting. 
These matrices are closely related to measurable quantities that are traditionally used by experimentalists 
to identify structures.

\subsection{Hamiltonian Matrix}
Emission and absorption  spectra  arise from transitions between discrete electronic energy levels.
Each element has its characteristic energetic levels and therefore atomic spectra can be used as elemental fingerprints. 
When atoms are assembled into structures the electronic states of the constituent atoms are modified 
depending on the arrangement of the atoms.
A computational analogue to electronic energy levels probed by various spectroscopic experiments 
are the Kohn-Sham energy eigenvalues, even though they do not represent the physical excitation energies. 
Since the Kohn-Sham Hamiltonian matrix depends only on the interatomic distances, the sorted 
Kohn-Sham eigenvalues are invariant to translations, rotations, reflections and permutations of atoms.


We examine fingerprints that are based on the occupied Kohn-Sham eigenvalues only 
as well as fingerprints that are based both on the occupied and unoccupied eigenvalues.
 The former were obtained from the self-consistent eigenvalues calculated in a large wavelet basis,~\cite{bigdft}
 whereas, for simplicity,  the latter were obtained 
from the non-self-consistent input guess eigenvalues calculated in a minimal Gaussian type atomic orbitals (GTO's) basis set
 for a charge density which is a superposition of atomic charge densities. 
Even though the length requirement of the coincidence axiom is violated in all cases, 
the configurational distances  $\Delta_{KS}(p,q)$ obtained from the occupied Kohn-Sham eigenvalues 
show a good correlation with the RMSD for the five test sets, see Fig.~\ref{fig:compare}. 
Fingerprint distances based on the vector $\bV_{GTO}$  do not much better correlate with the RMSD 
than fingerprint distances based on $\bV_{KS}$,  
even though the vector $\bV_{GTO}$  is in all cases longer than the vector $\bV_{KS}$ 
(e.g. $4n$ in case of the Si cluster i.e. two times longer) and hence the coincidence axiom is satisfied in all cases. 

Since different distances measure different kinds of dissimilarities,
it is not expected that they correlate perfectly for large distances. 
What is important is that all our metrics clearly allow to distinguish between distinct and non distinct configurations.

\subsection{Overlap Matrix}\label{sec:OM} 
A matrix which has similar properties as the Hamiltonian matrix is the overlap matrix expressed in terms of GTO's.
Contrary to the Hamiltonian, all elements of the overlap matrix can easily be calculated analytically (Appendix~\ref{apx:OM}).
In the simplest case where only uncontracted s-type GTO's are used, the resulting fingerprint consists of $n$ scalars.
Information about the radial distribution can be incorporated in the overlap matrix by adding p and d type GTO's. 
In this way the configurational fingerprint vector becomes also longer than $3 n-6$ and the coincidence axiom will be satisfied. 

If the fingerprint is used to calculate distances between our test set of local minima configurations, 
it turns out that adding p-type orbitals gives only a small improvement and adding additional d-type 
orbitals gives only a very marginal improvement.  This is related to the fact that it is very unlikely that 
two local minima lie on the hypersurface that leaves the fingerprint invariant. 
The distances obtained with this fingerprint, denoted by $\Delta_{OM}(p,q)$, correlate therefore  again well 
with the RMSD as shown in Fig.~\ref{fig:compare}. 
The width of the GTO's was in all our tests given by the covalent radius  
of the atom on which the GTO was centered.

\subsection{Hessian Matrix} 
The vibrational properties, which are frequently used experimentally to identify structures, are closely related to the Hessian matrix
which consists of the second order derivatives of the energy with respect to the atomic positions. The vibrational frequencies are up to 
a scaling factor related to the mass of the atoms equal to the square root of the eigenvalues of the Hessian matrix. 
This matrix also belongs to the class of matrices with the desired properties. 
Unfortunately the calculation of the Hessian is rather 
expensive in the context of a DFT calculation and can also be cumbersome with sophisticated force fields. 
We will therefore not further pursue approaches based on an Hessian which is calculated within the same high level method as the energy and forces.
It however turns out that eigenvalues or eigenvectors of the Hessian matrices which are derived from another cheaper 
potential such as the LJ potential give also  good fingerprints. 
This is shown in Fig.~\ref{fig:compare} for our six test systems
after the lengths were  scaled to the equilibrium bond-length of the LJ potential.

\subsection{Discussion} \label{discuss}
Various $n \times n$ matrices have been used previously to characterize molecular configurations. 
The contact matrix  from the graph theory
exhibits discontinuities when the atomic distances cross the 
cutoff radius. By introducing a smooth cutoff these discontinuities disappear
and the resulting matrix has been used as a fingerprinting tool in the SPRINT method.~\cite{sprint}
Presumably not only the contact matrix but also other matrices from spectral graph theory such as the Laplace
matrix could be used in a similar way. 
We did for instance not find significant differences in performance between the contact and Laplacian matrices.
We found however that fingerprints based on either of them 
are rather sensitive to the form of the smooth cutoff function.  
Tuning of the parameters of this cutoff function is therefore required to obtain good results.
In both cases, the resulting atomic fingerprints are real scalars which mostly contain information
about the number of nearest neighbours of each atom  
and might be insufficient to characterize the chemical environment of an atom.     
Better chemical environment descriptors can however be obtained by adding     
information about the radial distribution of the neighbours.~\cite{Steinhardt,Gabor12}

\begin{figure}[] \begin{center} 
\includegraphics[width=.18\textwidth]{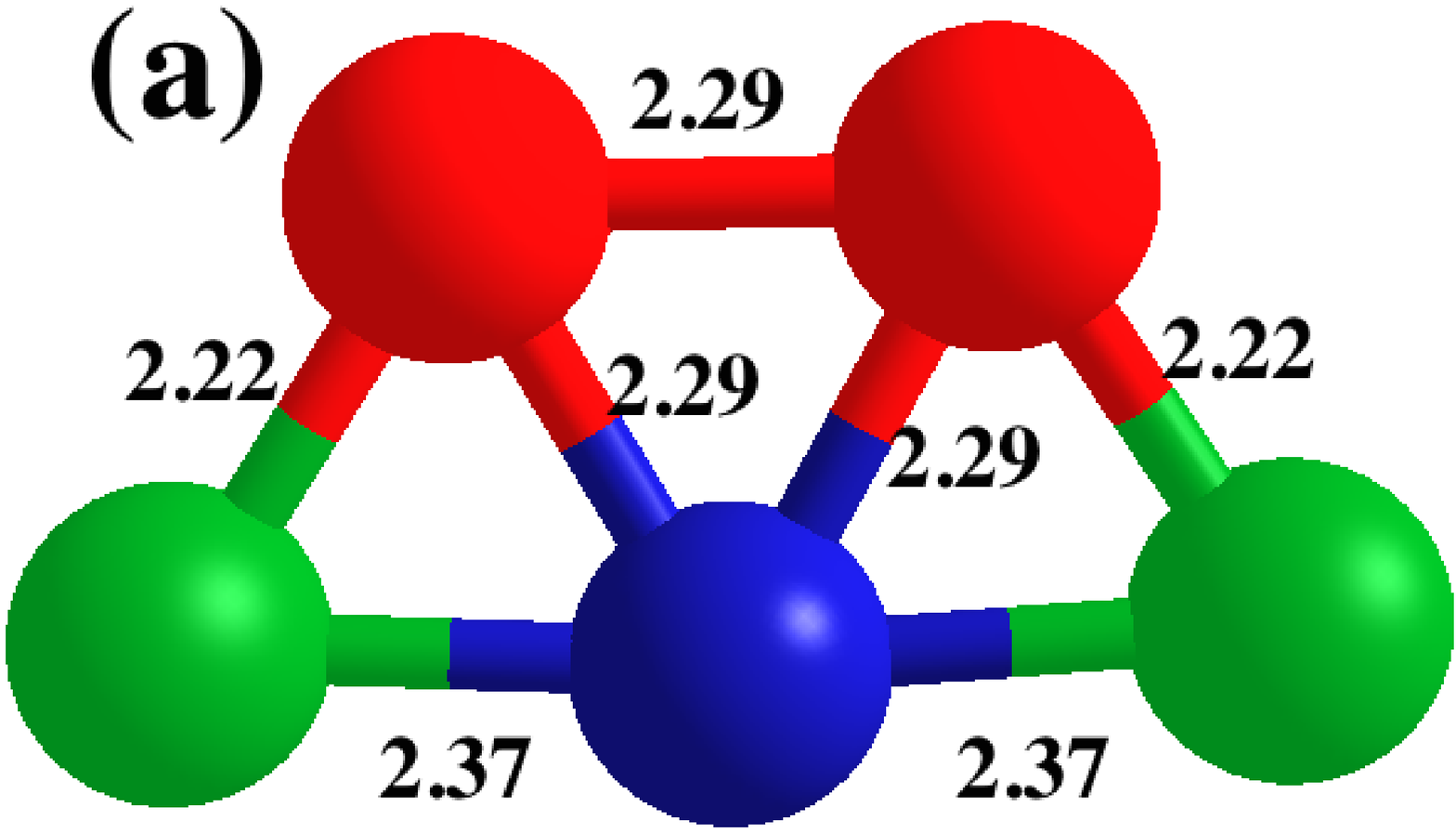}
\includegraphics[width=.14\textwidth]{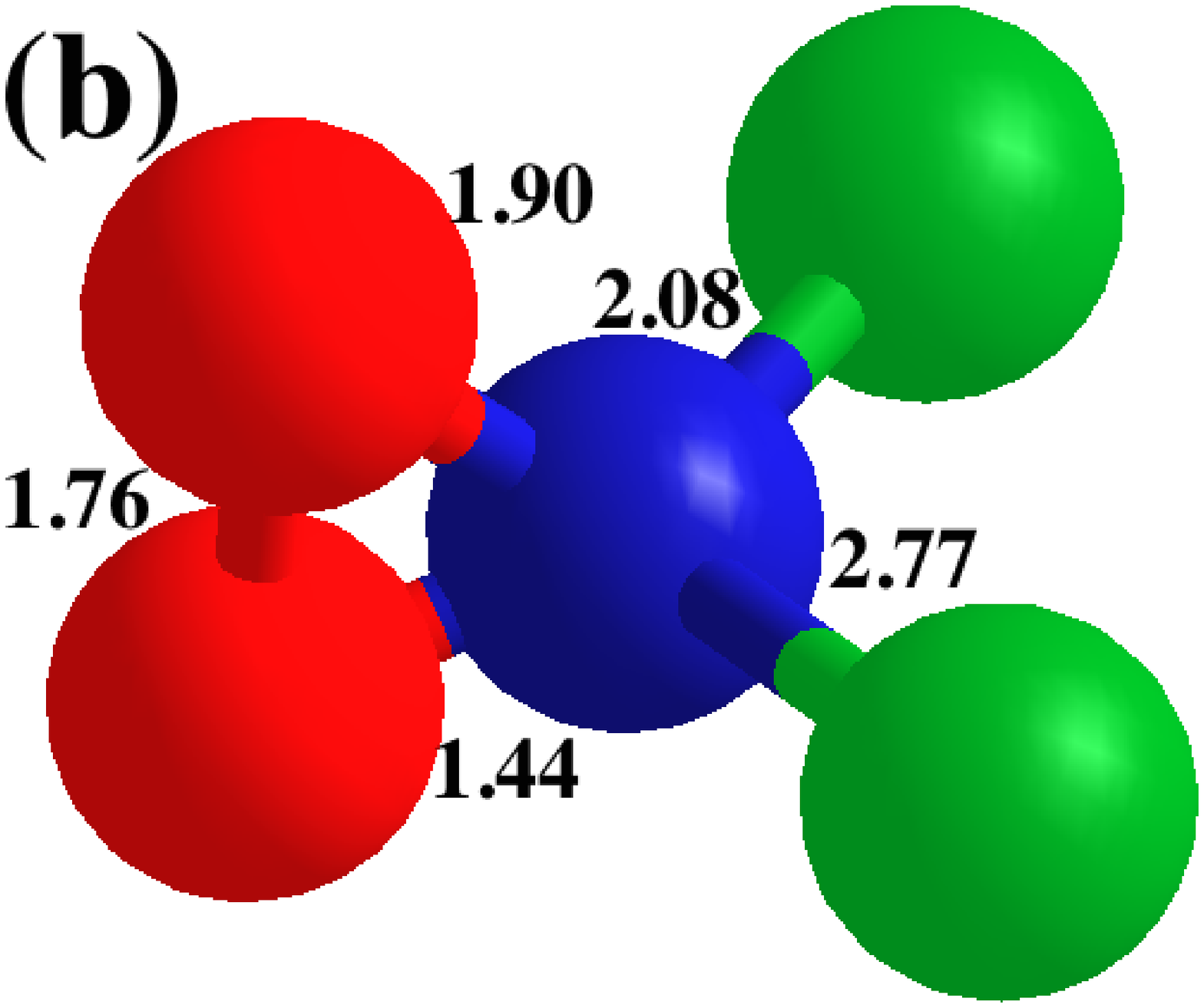}
\caption {\label{fig:sprint}  
Two distinct configurations of the Si$_5$ cluster with an identical set of SPRINT coordinates, 
i.e. 3.59 (green), 4.37 (red), 4.85 (blue), 
using the parameters given in the Supplementary Material of Ref.~\cite{sprint}.
The planar structure shown in (a) is a local minimum in DFT.
The numbers show the bond-lengths in \AA.
}
\end{center} \end{figure}

As discussed in before, a fingerprint of length $n$
does not satisfy the coincidence axiom and can thus fail to detect structural differences.
This has already been shown for the Coulomb matrix.~\cite{moussa}
We show in Fig.~\ref{fig:sprint} two distinct configurations of a Si$_5$ cluster 
which have identical sets of  SPRINT coordinates.
Note that the Si atoms with identical SPRINT coordinates in the configuration shown in  Fig.~\ref{fig:sprint}(b),
have very different environments. 
This shows that SPRINT, like any other $n\times n$ matrix-based fingerprint, 
fails to describe uniquely the entire  structure and/or the chemical environment of an atom.

\section{Conclusions}
In summary, we have shown that the RMSD, 
the most natural measure of dissimilarity between two configurations, 
satisfies the properties of a metric 
when it is obtained by a global minimization over all rotations and index permutations.
We have presented a Monte Carlo method to calculate the global minimal  RMSD which does not require to try out all
 possible index permutations and which is thus computationally feasible.
 At the same time we have introduced other metrics which  are  much cheaper to calculate 
because they do not require a structural superposition. 
Nevertheless they correlate in all our test cases very well with the RMSD. 
In contrast to numerous previously proposed fingerprints they satisfy the coincidence axiom 
and allow therefore to distinguish distinct from non-distinct configurations in a unique way. 
Within a DFT 
calculation the metric based on the Kohn-Sham eigenvalues is a good choice since the 
eigenvalues are a byproduct of any DFT calculation and thus no extra effort is required to obtain them. 
For the coincidence axiom to be satisfied, the number of bound eigenstates whose Kohn-Sham eigenvalues can be included in the 
fingerprint vector has however to be larger than $3n-6$. 
If Kohn-Sham eigenvalues are not available, the method based on the eigenvalues of the overlap matrix constructed from s and p orbitals is recommended, 
since it leads to matrices whose elements can be calculated analytically and because the fingerprint vector is long enough ($ 4 n$) to make the 
probability of a violation of the coincidence axiom vanishingly small. 
Even if the coincidence axiom is violated, it turns out in
practice that it is very rare that different physically reasonable metastable configurations give rise to identical
  fingerprints. For our test sets of low energy local minima configurations 
   metrics which violated the coincidence axiom therefore
  allowed nevertheless in all cases to distinguish between
 distinct and non-distinct configurations. In other applications where small movements away form metastable
 configurations lead to a change of physical properties,
  such as in force fields based on machine learning, a violation of the coincidence theorem can however not be
  tolerated. 
All the proposed variants of our approach are parameter free and no parameter tuning is therefore required.

\begin{acknowledgments}
We gratefully thank  D.G. Kanhere, S. De, R. Schneider and R. Ebrahimian for interesting  and helpful discussions.
This work has been supported by the Swiss National Science Foundation (SNF) and 
the Swiss National Center of Competence in Research (NCCR) on Nanoscale Science. 
Structures were visualized using  V\_Sim~\cite{vsim} and VESTA~\cite{vesta} packages.
Computing time was provided by the CSCS.
\end{acknowledgments}

\appendix

\setcounter{figure}{0}
\renewcommand{\thefigure}{A\arabic{figure}}

\section{Overlaps between  GTO's}\label{apx:OM}

\begin{figure} [] \begin{center}
\includegraphics[width=.3\textwidth]{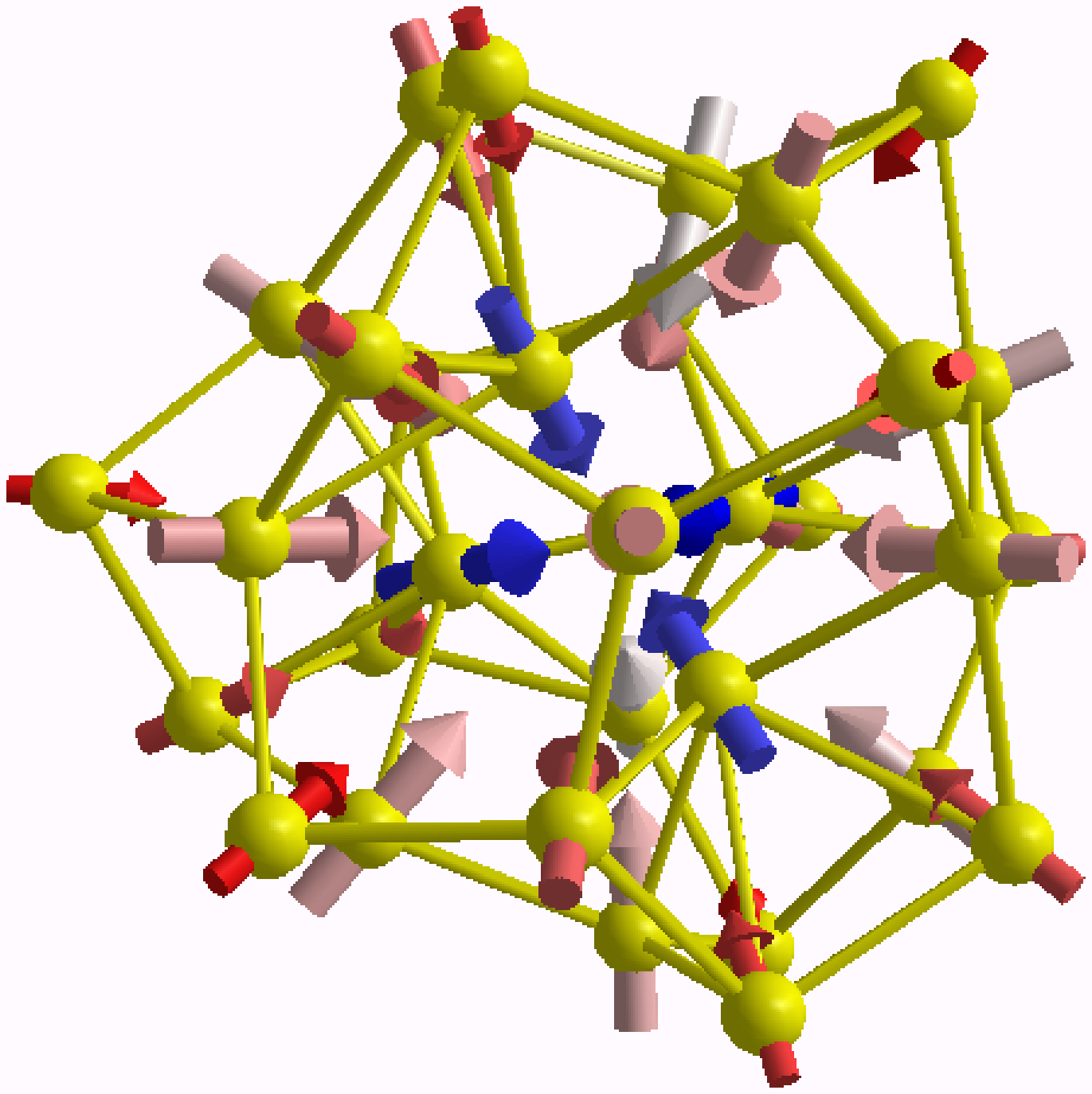}
\caption {\label{fig:atomicfp}
Description of atomic environments for a Si$_{32}$ cluster 
using the combined scalar and vectorial atomic fingerprints.
Each atomic fingerprint consists of a scalar and a vector  
which are the corresponding s and $(p_x,p_y,p_z)$ components 
of the principal eigenvector of the $4n\times 4n$ overlap matrix.
The color of the vectors indicates the value (red corresponds to small values  and blue to large values)
 of the scalar ($s$-type) fingerprint.
}
\end{center} \end{figure}

The normalized Gaussian type orbitals (GTO) centered at the atomic positions $\br_i$ 
in Cartesian coordinates are given by
\beq \nonumber
\phi_i^\bl (\br)= N_\bl
 (x-x_i)^{l_x}(y-y_j)^{l_y}(z-z_i)^{l_z}e^{-\alpha_i\|\br-\br_i\|^2}
\eeq
 where $\bl=(l_x,l_y,l_z)$ and $N_{\bl}$ is the normalization factor. 
Depending on the angular moment $L=l_x+l_y+l_z$ the functions are labeled as as s-type ($L$=0), p-type ($L$=1),
d-type ($L$=2) and so on.
We take the Gaussian width $\alpha_i$ inversely proportional to the square of 
the covalent radius of atom $i$ throughout this work.

The Gaussian product theorem says that the product of two Gaussian functions is again a Gaussian function.
Therefore the overlap integrals between a pair of GTO's, namely 
\beq \label{eq:ff}
\langle \phi_i^{\bl}|\phi_j^{\bl'} \rangle 
=\int d\br \phi_i^{\bl}(\br)\phi_j^{\bl'}(\br) 
\eeq
can be evaluated analytically. This gives the normalization factors as
\beq 
N_\bl(\alpha_i) = \frac 1{\sqrt{\langle \phi_i^{\bl}|\phi_i^{\bl} \rangle }}
&= \left({2\alpha_i}/{\pi}\right)^{3/4}\sqrt{n_{l_x} n_{l_y}n_{l_z}}, 
\nonumber \\ \nonumber
& n_{k} =\frac{(4\alpha_i)^{k}}{(2k-1)!!}
.\eeq 

All GTO's are recursively obtained by differentiating 
\[\phi_i^s(\br)=\big(\frac{2\alpha_i}{\pi}\big)^{3/4}e^{-\alpha_i\|\br-\br_i\|^2}\]
with respect to the Cartesian components of  $\br_i$. For instance
\[ \phi_i^{p_x}(\br)=2\sqrt {\alpha_i} (x-x_i) \phi_i^s(\br) \]
can also be expressed as 
\beq  \label{eq:phipx}
\phi_i^{p_x}(\br)=\frac 1 {\sqrt \alpha_i} \frac {\partial \phi_i^s(\br)}{\partial x_i}
.\eeq

The general formula for the overlap integrals, i.e. the elements of the overlap matrix,
is given, e.g., by Eq.~(3.5) in Ref.~\cite{clementi}
and can also be calculated from recursion relations.~\cite{Obara86}
For convenience, we restate the simplified relations for the special cases involving  
s and p-type GTO's all in terms of the basic quantity
\renewcommand{\O}{S_{ij}^{}}
\beq
\O=S_{ji}  =\Big(\frac{2\sqrt{\alpha_i \alpha_j}}{\alpha_i+ \alpha_j}\Big)^{ 3/2}
\exp\Big[ {\frac{-\alpha_i \alpha_j}{\alpha_i +\alpha_j} r_{ij}^2}\Big]
\eeq
where $r_{ij}=\|\br_i-\br_j\|$,
which is indeed the s-s overlap integral
\beq 
\langle \phi_i^s|\phi_j^s \rangle = S_{ij}
\nonumber 
\eeq
Using Eq.~(\ref{eq:phipx})  we obtain
\beq 
\langle \phi_i^{p_x}|\phi_j^s \rangle
&=& \frac {1} {\sqrt{\alpha_i}}  \frac{\partial \O}{\partial x_i}
\nonumber \\
&=&  
- \Big(\frac{2\sqrt{\alpha_i} \alpha_j}{\alpha_i+ \alpha_j}\Big) (x_i-x_j) \O
\eeq
and
\beq
 \langle \phi_i^{p_x}|\phi_j^{p_{x'}} \rangle 
= \Big(\frac{2 \sqrt {\alpha_i \alpha_j}}{\alpha_i+\alpha_j} \Big) \O 
& \\ \nonumber 
 \Big[\delta_{x,x'} &- \frac{2 {\alpha_i \alpha_j}}{\alpha_i+\alpha_j} (x_i-x_j)(x'_i-x'_j)\Big] 
\eeq
where $x,x'\in\{x,y,z\}$ 
and $\delta$  denotes the Kronecker delta.
The derivative of the basic quantity $S_{ij}$
 with respect to the atomic positions 
\beq \label{eq:dsijdxk}
\frac{\partial \O}{\partial x_k} 
&=& (\delta_{ik} - \delta_{jk} )  \Big(\frac{-2\alpha_i \alpha_j}{\alpha_i+ \alpha_j}\Big) (x_i-x_j) \O
\eeq
is required to calculate the derivative of the overlap matrix elements,
which in turn determine the derivative of its eigenvalues
 (see  Eq.~(\ref{LS}))
\beq
D_{\nu,x_{k}} \equiv \frac {\partial V_\nu} {\partial x_k}
= \langle \nu \big| \frac {\partial O}{\partial x_k} \big| \nu \rangle
,\eeq
where  the eigenvector $|\nu \rangle$ 
corresponds to the eigenvalue $V_\nu$ of the overlap matrix $O$.

\begin{figure}  []\begin{center}
\includegraphics[width=.4\textwidth]{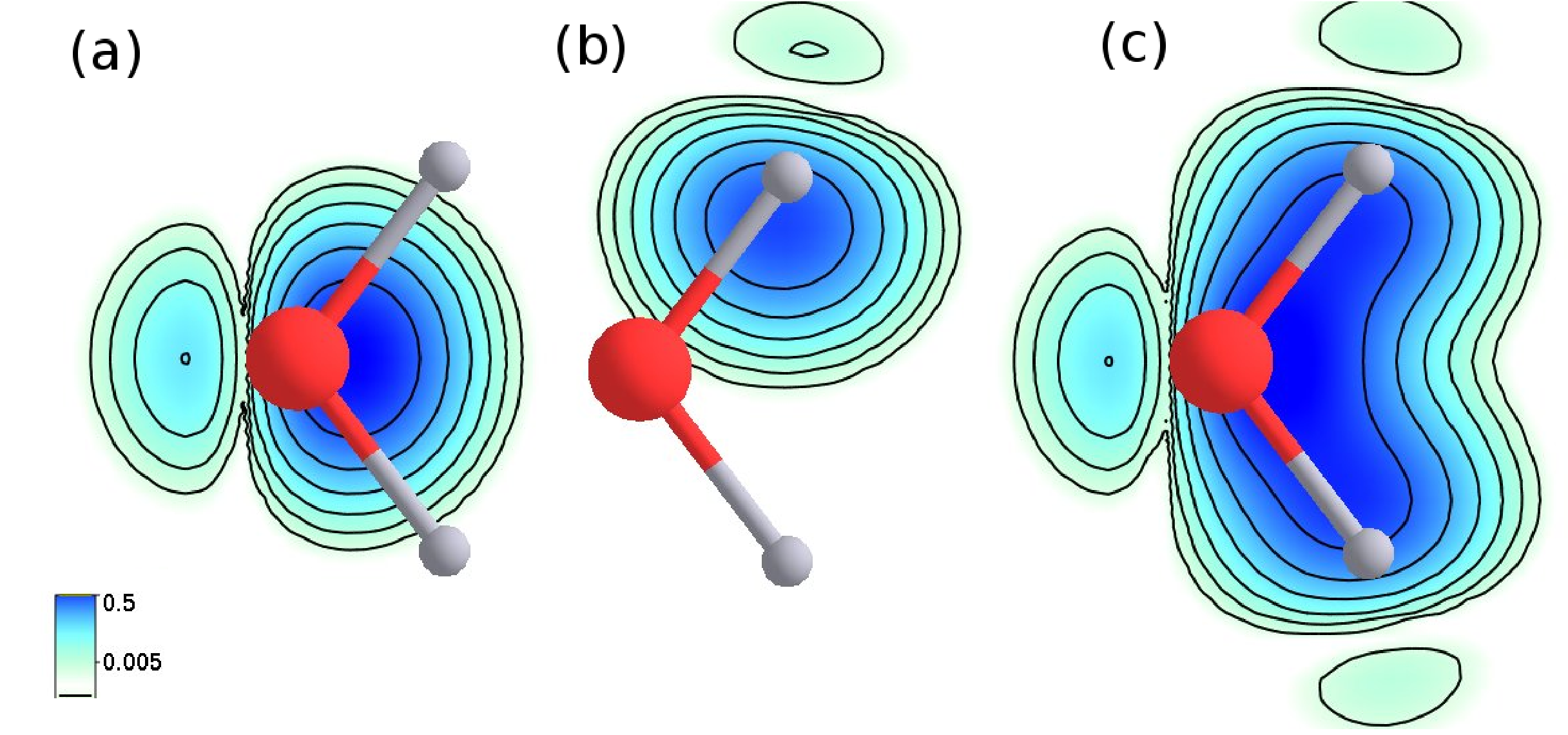}
\caption {\label{fig:isosurf}
Contributions from  an oxygen (a) or  hydrogen atom (b) to the total (c)
pseudo-charge density $|\psi(\br)|^2$ on the molecular plane for a water molecule.
The coefficients of the orbitals $\phi_i^\bl$ from which the pseudo-wavefunction $\psi$ is made,
 are the elements of the principal eigenvector of the overlap matrix constructed from s and p-type GTO's.}
\end{center} \end{figure}

Eigenvectors associated to small eigenvalues seem not to contain any useful information. 
We therefore use the principal eigenvector of the overlap matrix as an atomic fingerprint, see Fig.~\ref{fig:atomicfp}.
This vector gives the coefficients required to construct  the pseudo-orbital with the largest pseudo charge density. 
This charge density has similarities to a true charge density since it is large in regions between neighboring 
atoms where covalent bonding can occur (Fig~\ref{fig:isosurf}).

\section{Closed-form of superimposing rotation}\label{apx:Q}

\newcommand{\q}{\mathcal Q}
A quaternion $\q=(\q_0,\q_1,\q_2,\q_3)$ is an extension of the idea of complex numbers 
to  one real ($\q_0$) and three imaginary parts. 
According to the Euler's rotation theorem, a rotation in space 
which keeps one point on the rigid body (centroid in our case) fixed,
can be represented by four real numbers:
one for the rotation angle and three for the rotation axis
(we assume that the center of rotation is on the origin).
A  unit quaternion, i.e. $\|\q\|^2=\q_0^2+\q_1^2+\q_3^2+\q_4^2=1$,
 can represent conveniently this axis-angle couple as 
 \[\q=\Big(\cos\big(\frac \theta 2\big),\hat {\bf u}\sin\big(\frac \theta 2\big)  \Big)\]
where $\theta$ is the rotation angle around  the unit axis 
$\hat {\bf u}=a\hat {\bf i} + b\hat {\bf j}+c\hat {\bf k}$.
The corresponding orthogonal rotation matrix is
\begin{widetext}
\beq 
&& \bU=
\begin{bmatrix}
\q_0^2+\q_1^2-\q_2^2-\q_3^2&2\q_1\q_2-2\q_0\q_3        &2\q_1\q_3+2\q_0\q_2        \\
2\q_1\q_2+2\q_0\q_3        &\q_0^2-\q_1^2+\q_2^2-\q_3^2&2\q_2\q_3-2\q_0\q_1        \\
2\q_1\q_3-2\q_0\q_2        &2\q_2\q_3+2\q_0\q_1        &\q_0^2-\q_1^2-\q_2^2+\q_3^2\\
\end{bmatrix}
.\eeq
\end{widetext}

The optimum rotation $\bU$ which minimizes RMSD, indeed maximizes 
the correlation between $\bR^p$ and $\bR^q$, i.e. the atomic Cartesian coordinates
with respect to the common center of mass.
Based on quaternions,~\cite{Horn}
the optimum $\bU$ is given by $\q$ which is identical to the principal eigenvector of 
the 4$\times$4 symmetric, traceless matrix
\begin{widetext}
\beq
\mathcal F=
\begin{bmatrix}
  \mathcal R_{xx}+\mathcal R_{yy}+\mathcal R_{zz} & \mathcal R_{yz}-\mathcal R_{zy} & \mathcal R_{zx}-\mathcal R_{xz} & \mathcal R_{xy}-\mathcal R_{yx} \\
  \mathcal R_{yz}-\mathcal R_{zy} & \mathcal R_{xx}-\mathcal R_{yy}-\mathcal R_{zz} & \mathcal R_{xy}+\mathcal R_{yx} & \mathcal R_{xz}+\mathcal R_{zx} \\
  \mathcal R_{zx}-\mathcal R_{xz} & \mathcal R_{xy}+\mathcal R_{yx} &-\mathcal R_{xx}+\mathcal R_{yy}-\mathcal R_{zz} & \mathcal R_{yz}+\mathcal R_{zy} \\
  \mathcal R_{xy}-\mathcal R_{yx} & \mathcal R_{xz}+\mathcal R_{zx} & \mathcal R_{yz}+\mathcal R_{zy} &-\mathcal R_{xx}-\mathcal R_{yy}+\mathcal R_{zz}  
 \end{bmatrix} 
\eeq
\end{widetext}
where $\mathcal R$ is the correlation matrix whose elements are $\mathcal R_{xy}=\sum_i^n x_i^p y_i^q$ and so.
Eq.~(\ref{eq:RMSDl}) 
is then given by
\beq  
 RMSD(p,q)=\sqrt{\frac 1 n \Big( \|\bR^p\|^2 + \|\bR^q\|^2-2\lambda^{*} \Big) }
\eeq
where $\lambda^{*}$ is the largest eigenvalue of $\mathcal F$.


%

\end{document}